\documentclass[aps,pre,epsfig,floats,twocolumn,superscriptaddress,amssymb,amsmath,floatfix,showpacs]{revtex4}
\usepackage{graphicx}
\usepackage{bm}  
\usepackage{amssymb}
\usepackage{color}
\usepackage{amscd}

\begin{document}
\title{Hydrodynamic Interactions of Spherical Particles in A Fluid Confined by A Rough And No-Slip Wall}

\author{Somaye Hosseini Rad}
\affiliation{Department of Physics, Zanjan University, Zanjan 313, Iran}

\author{Ali Najafi\footnote{To whom correspondence should be addressed.}
}
\email{najafi@znu.ac.ir} 
\affiliation{Department of Physics,
Zanjan University, Zanjan 313, Iran}

\date{\today}

\begin{abstract}
In this article we develop a theoretical framework to study the hydrodynamic interactions in the presence of a non-flat and no-slip boundary. 
We calculate the influence of a small amplitude and sinusoidal deformations of a boundary wall in the self mobility and the two body hydrodynamic interactions for spherical particles. 
We show that the surface roughness enhances the self mobility of a sphere in a way that, for motion in front of a local hump of the  surface, the mobility strength  decreases while it increases for the motion above a local deep of the rough surface. 
The influence of the surface roughness in the two body hydrodynamic interactions is also analyzed numerically. 
\end{abstract}
\maketitle
\section{Introduction}
Hydrodynamic interaction of colloidal particles in confined geometries is an important problem in low Reynolds fluid dynamics~\cite{happel}.
Fluid motions confined by one or two parallel flat planes are interesting examples with either analytical or numerical known solutions~\cite{blake,twoplate}. 
The solution to these problems are involved in soft matter related phenomena and also microfluidic experiments~\cite{micro,microhyd,zurita}. 
In soft matter systems, swimming motion in a geometrically confined environment is a subject of
growing interest~\cite{lowen2,nanorod-wall,raminconfined,bacteriasurface}, apparently better understanding of these systems requires a good knowledge of the hydrodynamic interactions in confined geometries.
In microfluidic applications, a better control of the processes requires the prediction of the hydrodynamic effects due to the walls.

Confining wall  can generate nontrivial effects. For example,  
in a very simple system composed of a single sphere near a wall, the mobility parallel to the wall is always larger than
the perpendicular mobility~\cite{happel} {which} has been
verified experimentally~\cite{pralle,dufresne}. As other examples for the effects due to the boundaries, 
we address the experiments,
showing that
microorganisms, e.g. E. Coli~\cite{chasb}, bull
spermatozoa~\cite{chasb-ref6}, swimming in confined geometries are
attracted by surfaces. 
In this article we concentrate on the effects due to the roughness of the confining walls. We will consider a rough wall that 
confines the 
fluid flow at low Reynolds number, and ask the following question, does the long wavelength roughness of the wall have any important 
influence 
on the one or two body hydrodynamic interactions? We use a perturbation method and investigate the case of a small amplitude and 
regular roughness on an infinite wall. We show that the roughness has important contributions in the hydrodynamic interactions.

Another challenging issue in the low Reynolds, quiescent fluid dynamics, is the validity of no-slip boundary condition. 
This is certainly 
important where the nano structure of the surface is involved. There are experimental and theoretical 
works, investigating the influence of the nano-roughness of the boundaries and justify the validity of the so-slip boundary 
conditions~\cite{HS-noslip,rough-boundary}. 
Here we would like to stress the fact that, justification of the no-slip boundary condition, 
necessarily needs analytical results for rough surfaces taking into account the no-slip boundary condition.

The rest of this article is organized as follows: In Sec. II, we present a short review on Stokes flow and 
introduce the hydrodynamic interactions. Section III is
devoted to the hydrodynamic effects of a rigid and flat wall. The effects of a general roughness is presented in sec. IV. 
An example of a sinusoidal roughness is considered in Sec. V. Concluding remarks are presented in sec. VI.

\section{Stokes flow and Hydrodynamic Interactions}
\begin{figure}[t]
\includegraphics[width=.85\columnwidth]{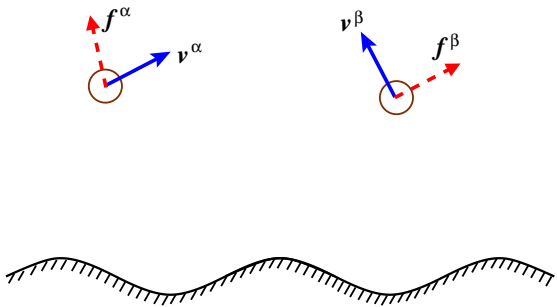}
\caption{(color online). Two spherical particle is moving in a low Reynolds flow. The fluid flow is bounded by a rough, rigid and no-slip 
boundary. Velocities and hydrodynamic forces acting on particles are shown in this picture.}
\label{fig1}
\end{figure}

To study the fluid motion for colloidal particles suspended in a fluid medium we define the Reynolds number as the ratio between the characteristic transport time scale due to diffusion and the convection time over a length $L$. Denoting the fluid density by $\rho$, viscosity by $\eta$, typical velocity by $U$ and also the linear size of the particles by $a$, we see that the Reynolds number is $Re=\frac{Ua\rho}{\eta}$. In a wide variety of phenomena occurring in the motion of micron scale particles, the Reynolds number is very low. For these phenomena we can consider the limit of zero Reynolds number. At zero Reynolds number the fluid dynamics is expressed by Stokes equation. Denoting the fluid velocity and pressure fields by ${\bf u}({\bf x})$ and $P({\bf x})$, the Stokes and continuity equations 
for an incompressible flow can be written as:
\begin{equation}
\eta\nabla^2{\bf u}({\bf x})-\nabla P({\bf x})={\bf f}^{ext},~~\nabla\cdot{\bf u}({\bf x})=0,
\label{stokes}
\end{equation}
where ${\bf f}^{ext}$, denotes the density of external body force acting on the fluid. 
The fluid velocity field is subject to no-slip boundary condition on the surface of solid boundaries. 
We would like to consider the motion of $N$ colloidal particles suspended is a low Reynolds flow (see Fig. \ref{fig1}). Recalling the fact that the governing equations are linear with respect to the velocity profile,  and applying the no-slip boundary condition on the surface of rigid 
particles, we can eliminate the fluid velocity and obtain a set of linear equations relating the particle velocities to the hydrodynamic forces acting on them. Denoting the velocity of $m$'th particle by ${\bf v}^{m}$ and the corresponding hydrodynamic force by 
${\bf f}^{m}$, we can write the following equation:
\begin{equation}
{\bf v}_{i}^{m}=\sum_{n=1}^{N}\sum_{j=1}^{3}D_{ij}^{mn}\times{\bf f}_{j}^{n}.
\end{equation}
Coefficients $D_{ij}^{mn}$ are know as the elements of hydrodynamic interaction.
Quite similar to the Onsager relations in thermodynamics and based on general symmetry arguments and microscopic reversibility, 
the hydrodynamic tensor $D_{ij}^{mn}$ is symmetric with respect to its up or down indices \cite{reichel,dhont}. For special case of 
$2$ particles (denoted by $\alpha$ and $\beta$), and for convenience we define the self mobility 
tensor as the response of a particle to the force acting on it by $\mu_{ij}=D_{ij}^{\alpha\alpha}$. And also the hydrodynamic response of 
$\alpha$'th particle to the force acting on $\beta$'th particle is shown by: $M_{ij}=D_{ij}^{\alpha\beta}=D_{ij}^{\beta\alpha}$. 
Different components of the hydrodynamic interaction  tensor depend on the size and the relative configurations of the particles.
Calculating the hydrodynamic interaction tensor, is an important problems in low Reynolds hydrodynamics. Here we present an analytic  
method that is useful for very small and spherical particles.  A very small 
sphere can be considered as a point force, a source term in the Stokes equation. Denoting the point force by ${\bf f}({\bf x})={\bf
b}~\delta({\bf x}-{\bf x}_0)$, the
velocity field of the point force, can be written as~\cite{Oseen,pozrikidis}: 
\begin{equation}
{\bf u}_i({\bf
x})=\sum_{j=1}^{3}{G_{ij}}({\bf x},{\bf x}_0)\times{\bf b}_{j},
\label{greendefinition}
\end{equation}
and correspondingly the associated pressure field is given by: 
$P({\bf x})={\bf \Pi}({\bf x},{\bf x}_0)\cdot{\bf b}$. Here $G_{ij}({\bf x},{\bf x}_0)$ is the Green's function of the 
Stokes differential equation.

Solving the Stokes equations and having the associated Green's function for any required geometry, we can use
Fax\'{e}n's theorem {for spherical objects with radius $a$} to
express the mobility tensor in terms of the Green's function of
Stokes equation~\cite{happel,Kim}:
\begin{equation}
D_{ij}^{\alpha\beta}({\bf x}^{\alpha},{\bf x}^{\beta})=(1+\frac{a^2}{6}\nabla^{2}_{{\bf
x}^\alpha})(1+\frac{a^2}{6}\nabla^{2}_{{\bf x}^\beta})G_{ij}({\bf
x}^\alpha,{\bf x}^\beta).
\label{faxen}
\end{equation}
For very small spheres with radius $a$ smaller than the typical distance
between spheres, then $D_{ij}^{\alpha\beta}({\bf x}^{\alpha},{\bf x}^{\beta})=
G_{ij}^{F}({\bf x}^\alpha,{\bf x}^\beta)$ up to second order in $a$~\cite{infinite}. 

In the rest of this article, we use the above method and presents the results for unconfined ($U$), confined with a flat wall ($F$) and 
confined with a rough wall ($R$). The effects due to roughness is considered for a very small amplitude roughness.
\section{Hydrodynamics Near a Flat and No-Slip Wall}
We first consider the case of a fluid flow that is bounded by a rigid, flat ($F$) and no-slip plane located at $z=0$. The solution to the problem of a point force in the presence of a flat wall is given by Blake \cite{blake,pozrikidis}, who has used the image method to construct the solutions.
The flat wall Green's function for the upper half space ($z>0$), is given by:
\begin{eqnarray}
 G_{ij}^{F}({\bf x},{\bf x}_0)&=&
G_{ij}^{U}({\bf x},{\bf x}_0)-G_{ij}^{U}({\bf x},{\bf x}^{im}_{0})\nonumber\\
&+&G_{ij}^{D}({\bf x},{\bf x}^{im}_{0})-G_{ij}^{SD}({\bf x},{\bf x}_{0}^{im}),
\end{eqnarray}
where the  Green's function for an unbounded ($U$) fluid flow $G_{ij}^{U}$, that is called stokeslet is given by:
\begin{equation}
 G_{ij}^{U}({\bf x},{\bf x}_0)=\frac{1}{8\pi\eta}\left(\frac{\delta_{ij}}{r}+\frac{r_ir_j}{r^3}\right),
\end{equation}
here ${\bf r}={\bf x}-{\bf x}_0$, and ${\bf x}^{im}_{0}={\bf x}_0-2\left({\hat z}\cdot{\bf x}_0\right){\hat z}$ is the image position of the point force with respect to the flat wall. Defining a 
new vector ${\bf R}={\bf x}-{\bf x}^{im}_{0}$, we can express the potential dipole field $G_{ij}^{D}$, and stokeslet dipole $G_{ij}^{SD}$ as:
\begin{equation}
 G_{ij}^{D}({\bf x},{\bf x}^{im}_{0})=\frac{2}{8\pi\eta}z_{0}^{2}(1-2\delta_{jz})\frac{\partial}{\partial R_j}\left(\frac{R_i}{R^3}\right),
\end{equation}
\begin{equation}
 G_{ij}^{SD}({\bf x},{\bf x}^{im}_{0})=2z_{0}(1-2\delta_{jz})\frac{\partial}{\partial R_j}G_{iz}^{U}
({\bf x},{\bf x}^{im}_{0}),
\end{equation}
where $z={\hat z}.{\bf x}$. The pressure field for the flow that is bounded by a flat wall is given by:
\begin{eqnarray}
{\bf \Pi}^{F}({\bf x},{\bf x}_0)&=&{\bf \Pi}^{U}({\bf x},{\bf x}_{0})+{\bf \Pi}^{U}({\bf x},{\bf x}^{im}_{0})\nonumber\\
&-&2z_{0}\frac{\partial}{\partial{\bf R}}\left({\hat z}\cdot{\bf \Pi}^{U}({\bf x},{\bf x}^{im}_{0})\right),
\end{eqnarray}
where the pressure field associated to a point force in an unbounded space is given by:
\begin{equation}
{\bf \Pi}^{U}({\bf x},{\bf x}_0)=\frac{2}{8\pi}~\frac{\left({\bf x}-{\bf x}_0\right)}{|{\bf x}-{\bf x}_0|^3},
\end{equation}
Having in hand all the above information, we can calculate the hydrodynamic interactions in the presence of a flat no-slip wall.

As an example and using the above formalism we present the results for the self mobility of a very small sphere moving above a flat 
and no-slip wall. Denoting the perpendicular distance between the sphere and wall by $H$, and assuming that $H$ is 
greater than the sphere radius $a$, the components of self mobility tensor are~\cite{happel,Kim}
\begin{eqnarray}
\mu^{F}_{xx}&=&\mu_0(1-\frac{9}{16}\frac{a}{H}+{\cal O}(\frac{a}{H})^2),\nonumber\\
\mu^{F}_{zz}&=&\mu_0(1-\frac{9}{8}\frac{a}{H}+{\cal O}(\frac{a}{H})^2),
\label{selfmobility}
\end{eqnarray}
where $\mu_0={1}/{(6\pi\eta a)}$ is the self mobility of a spherical particle moving in an unbounded fluid. 
As an important and non-trivial result, the above equations show that, 
the self mobility of the sphere in the direction parallel to the wall is always larger than
the perpendicular direction~\cite{happel}. This effect has been
verified experimentally~\cite{pralle}. Calculations show that, up to ${\cal O}(\frac{a}{H})^2$, other components of the self mobility tensor are zero.
\section{Effects Due to a Rough and No-Slip Wall}
The aim of this article is to express analytical expressions for hydrodynamic interactions in a semi infinite flow, bounded by a 
rough ($R$), rigid and no-slip wall. Fig. \ref{fig1} shows the schematic view of a rough plane that bounds the fluid flow. 
Two spherical particles located at positions ${\bf x}^{\alpha}$ and ${\bf x}^{\beta}$ are moving in the fluid. 
The position vector for the points on the bounding wall is considered as: ${\bf x}_s=(x,y,h(x,y))$. This kind of parametrization allows us 
to obtain the results of a flat wall by considering the limiting case of $h(x,y)=0$. 
For later use we define ${\bf x}^{0}_{s}=(x,y,0)$, that is 
denoting the location of the points of a flat wall located at $z=0$.

Applying the no-slip boundary condition on the rough surface ($R$), 
we expect that the influence of the roughness will produce non-trivial effects. 
To study the hydrodynamic effects of a rigid and rough wall with small amplitude roughness, we can construct a perturbation 
expansion. Introducing a small dimensionless parameter $\varepsilon={h_0}/{z}$, where $h_0$ is the typical amplitude for height 
fluctuations and $z$ measures the distance of particles from the wall, we expand all quantities in powers of $\varepsilon$. 
As explained before, hydrodynamic interaction between small spherical particles 
can be obtained by solving the velocity field of a point force for the required geometry. 
Here the solutions to the Stokes-Green's equation should be obtained by applying the required boundary condition. 
The velocity field of a point 
force satisfies the Stokes equation (Eq. \ref{stokes}) and is subject to to the following boundary condition:
\begin{equation}
{\bf u}({\bf x}_s)=0.
\end{equation}
Expanding the corresponding velocity and pressure field of a point force in the presence of a rough wall 
in powers of $\varepsilon$, we 
can write:
\begin{eqnarray}
{\bf u}^{R}({\bf x})&=&{\bf u}^{(0)}({\bf x})+{\bf u}^{(1)}({\bf x})+{\bf u}^{(2)}({\bf x})+{\cal O}(\varepsilon^3),\nonumber\\
P^{R}({\bf x})&=& P^{(0)}({\bf x})
+P^{(1)}({\bf x})+P^{(2)}({\bf x})+{\cal O}(\varepsilon^3),
\end{eqnarray}
The zeroth order term are the velocity and pressure field of point force in the presence of a flat($F$) and no-slip wall. In this case we will have:
\begin{equation}
{\bf u}^{(0)}({\bf x})={\bf u}^{F}({\bf x}),~~{\bf P}^{(0)}({\bf x})={\bf P}^{F}({\bf x}).
\end{equation}
Higher order corrections due to the wall roughness, can be obtained by noting that different order of the velocity and 
pressure fields are satisfied the following differential equation:
\begin{equation}
\eta\nabla^2{\bf u}^{(n)}({\bf x})-\nabla P^{(n)}({\bf x})=0,~~\nabla\cdot{\bf u}^{(n)}({\bf x})=0, 
\end{equation}
where the boundary conditions for the first and second orders are given explicitly by:
\begin{eqnarray}
{\bf u}^{(1)}({\bf x}_{s}^{0})&=&-h({\bf x}_{s}^{0})\frac{\partial {\bf u}^{(0)}({\bf x}_{s}^{0})}{\partial z}.\nonumber\\
{\bf u}^{(2)}({\bf x}_{s}^{0})&=&-h({\bf x}_{s}^{0})\frac{\partial {\bf u}^{(1)}({\bf x}_{s}^{0})}{\partial z}-\frac{h^2({\bf x}_{s}^{0})}{2}\frac{\partial^2 {\bf u}^{(0)}({\bf x}_{s}^{0})}{\partial^2 z}.\nonumber\\
\end{eqnarray}
As one can see the velocity field at order $(n)$, is related to the local value of the derivatives of the velocity filed at order of 
$(n-1)$, in the position of flat wall. 
Using the well known integral representation for the flow field of Stokes equation, we can write the following integral solution~\cite{pozrikidis}:
\begin{equation}
{\bf u}_{j}^{(n)}({\bf x})=\frac{1}{8\pi}\int {\bf u}_{i}^{(n)}({\bf x'}){\bf T}_{ijk}({\bf x,x'})dS_k({\bf x'}), 
\label{velocitycorrection}
\end{equation}
where the integration is carried out on a closed surface composed of an infinite plane at $z=0$. The surface is closed at the upper half space and $d{\bf S}({\bf x}')$ is the area element on this surface in the inward direction (here is ${\hat z}$). The stress 
tensor ${\bf T}_{ijk}({\bf x},{\bf x}')$  is given by:
\begin{eqnarray}
{\bf T}_{ijk}({\bf x},{\bf x}')&=&-8\pi\delta_{ik}\Pi^{F}_{j}({\bf x},{\bf x}')+8\pi\eta\frac{\partial}{\partial x_k} G^{F}_{ij}({\bf x},{\bf x}')
\nonumber\\
&+&8\pi\eta\frac{\partial}{\partial x_i}G^{F}_{kj}({\bf x},{\bf x}').
\end{eqnarray}
Now we can expand the hydrodynamic interaction tensor moving in a fluid bounded by a rough wall, 
in terms of small parameter $\varepsilon$ as:
\begin{equation}
{\bf D}^{R}={\bf D}^{(0)}+{\bf D}^{(1)}+{\bf D}^{(2)}+{\cal O}(\varepsilon^3),
\end{equation}
where the zeroth order is the results of a sphere moving in a medium bounded by a flat ($F$) and no-slip wall:
\begin{equation}
D^{(0)}_{ij}=D^{F}_{ij}\approx G_{ij}^{F}.
\end{equation}
Using Eqs. (\ref{greendefinition}), (\ref{faxen}), (\ref{velocitycorrection}) and considering the explicit form of the stress tensor ${\bf T}_{ijk}({\bf x}_{\beta},{\bf x}_{0}^{s})$, we can arrive at the following 
expressions for the first and  second order corrections to the mobility tensor due to the roughness of the wall:
\begin{widetext}
\begin{eqnarray}
8\pi\eta\times D^{(1)}({\bf x}_{\alpha},{\bf
x}_{\beta})&=&-\frac{1}{8\pi}\int D'^{(0)}({\bf x}_{\alpha},{\bf
x}_{s}^{0})D'^{(0)}({\bf x}_{\beta},{\bf
x}_{s}^{0})h({\bf x}_{s}^{0})d^2{\bf x}_{s}^{0},
\end{eqnarray}
\begin{eqnarray}
256\pi^2\eta\times D^{(2)}({\bf x}_{\alpha},{\bf
x}_{\beta})&=&\frac{1}{2\pi}\int\int D'^{(0)}({\bf x}_{\alpha},{\bf
x}_{s}^{0})D'^{(0)}({\bf x}_{\beta},{\bf
x'}_{s}^{0})D''^{(0)}({\bf x}_{s}^{0},{\bf
x'}_{s}^{0})h({\bf x}_{s}^{0})h({\bf x'}_{s}^{0})d^2{\bf x}_{s}^{0}d^2{\bf x'}_{s}^{0}\nonumber\\
&~~~&-
\int\Big( E^{(0)}({\bf x}_{\alpha},{\bf
x}_{s}^{0}){\ddot D}^{(0)}({\bf x}_{\beta},{\bf
x}_{s}^{0})~+~(\alpha\leftrightarrow\beta) \Big)h^2({\bf x}_{s}^{0})d^2{\bf x}_{s}^{0},\nonumber\\
\end{eqnarray}
\end{widetext}
where 
\begin{eqnarray}
D'^{(0)}({\bf x}_{\beta},{\bf x}_{s}^{0})&=&\frac{\partial}{\partial z}D^{(0)}({\bf x}_{\beta},{\bf x})\mid_{{\bf x}\rightarrow{\bf x}_{s}^{0}}
\nonumber\\
D''^{(0)}({\bf x}_{s}^{0},{\bf x'}_{s}^{0})&=&\frac{\partial^2}{\partial z\partial z'}D^{(0)}({\bf x},{\bf x'})\mid_{({\bf x},{\bf x'})\rightarrow({\bf x}_{s}^{0},{\bf x'}_{s}^{0})}\nonumber\\
{\ddot D}^{(0)}({\bf x}_{\beta},{\bf x}_{s}^{0})&=&\frac{\partial^2}{\partial z^2}D^{(0)}({\bf x}_{\beta},{\bf x})\mid_{{\bf x}\rightarrow{\bf x}_{s}^{0}},
\end{eqnarray}
and $E^{(0)}_{ij}=\Pi^{(F)}_{ij}\delta_{i=z}+D'^{(0)}_{ij}\delta_{i\neq z}$.
As one can see, the mobility tensor is symmetric: $D({\bf x}_{\alpha},{\bf x}_{\beta})=D({\bf x}_{\beta},{\bf x}_{\alpha})$.
\begin{figure}[t]
\includegraphics[width=.85\columnwidth]{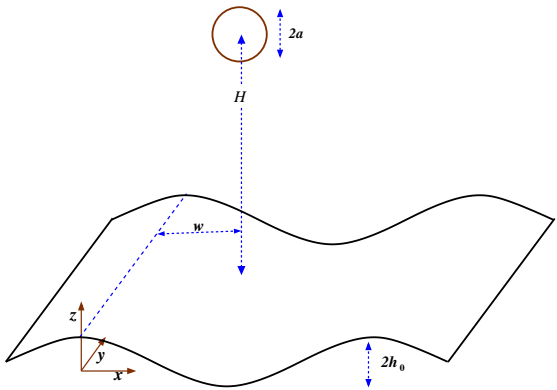}
\caption{(color online). A spherical particle is moving in a confined fluid flow. The roughness of the boundary is assumed to be a 
sinusoidal pattern with wavelength $\lambda$ in the $x$ direction. The boundary is assumed to be invariant under translation along the $y$ 
direction. Two parameters, $H$ the average vertical distance from the wall, and $w$ the vertical deviation from a local hump of the surface 
roughness, characterize the sphere's position.}
\label{fig2}
\end{figure}

In the following section we 
choose a very special form of the wall deformation pattern and investigate the hydrodynamic interactions. 
\section{Sinusoidal Roughness}
As a special example we assume that the wall roughness  is a  sinusoidal deformations of a flat wall located at $z=0$. The height profile of the rough plane, in the Monge representation, is given by the following equation:
\begin{equation}
z=h(x,y)=h_0\cos(qx+\phi),
\end{equation}
where $h_0$ is the amplitude of the roughness, and the wave vector of the roughness along the $x$ axis is denoted by $q$. This special type 
of roughness allows us to study any general roughness that is invariant under translation  along $y$ direction. 

As a special feature of the hydrodynamic interaction we concentrate on the self mobility of a spherical particle moving in a medium 
bounded by a rough and no-slip wall. Fig. \ref{fig2}, shows the schematic view of a sphere with size $a$, moving near a rough wall. 
Perpendicular projection of the sphere position into the $z=0$ plane, is deviated from the wall hump by a distance $w$. 
Denoting the wave length of the wall roughness by $\lambda=\frac{2\pi}{q}$, and the average height 
of the particle on the wall by $H$, we define a dimensionless parameter by $\gamma=2\pi\frac{H}{\lambda}$. Now following the method developed 
in the preceding sections, the first order corrections to the different components of the self mobility are given by:
\begin{eqnarray}
\mu_{xx}^{(1)}&=&\frac{1}{8\pi\eta H}\frac{h_0}{H}\cos(\phi)\Big[-\frac{9}{32}\gamma^2 K_2(\gamma)\nonumber\\
&~~~&+36\gamma^2\int_{0}^{\infty}\frac{s^5J_2(s)}{(s^2+\gamma^2)^5}ds\Big],\nonumber\\
\mu_{yy}^{(1)}&=&\frac{1}{8\pi\eta H}\frac{h_0}{H}\cos(\phi)\Big[-\frac{9}{32}\gamma^2 K_2(\gamma)\Big],\nonumber\\
\mu_{zz}^{(1)}&=&0,\nonumber\\
\end{eqnarray}
and the second order corrections are given by:
\begin{eqnarray}
\mu_{xx}^{(2)}&=&\frac{1}{8\pi\eta H}\frac{h_{0}^{2}}{H^2}\Big\{\frac{3}{32}+\cos(2\phi)\Big[\frac{15}{32}\gamma^3 K_3(2\gamma) \nonumber \\
&~~~&-\frac{9}{8}\gamma^4 K_0(2\gamma)+\frac{3}{8}\gamma^5 K_1(2\gamma)\nonumber \\
&~~~&-\frac{9}{8}\gamma^3K_1(2\gamma)-9\gamma^6\int_{0}^{\infty}\frac{s^3J_0(2s)}{(s^2+\gamma^2)^5}ds\Big] \nonumber \\
&~~~&+36\cos^2(\phi)e^{-\gamma}(1-\gamma)\Big[I_2(\gamma/2) K_2(\gamma/2)\nonumber \\
&~~~&-\gamma^4\int_{0}^{\infty}\frac{J_2(s)}{(s^2+\gamma^2)^{5/2}}ds \Big]\Big\},\nonumber \\
\mu_{yy}^{(2)}&=&\frac{1}{8\pi\eta H}\frac{h_{0}^{2}}{H^2}\Big\{\frac{3}{32}+\frac{15}{32}\cos(2\phi)\gamma^3 K_3(2\gamma)\nonumber\\
&~~~&+36\cos^2(\phi)e^{-\gamma}I_2(\gamma/2)K_2(\gamma/2)\Big\},\nonumber\\
\mu_{zz}^{(2)}&=&\frac{1}{8\pi\eta H}\frac{h_{0}^{2}}{H^2}\Big\{\frac{9}{8}+\frac{9}{8}\cos(2\phi)\Big[-\gamma^4K_4(2\gamma)\nonumber\\
&~~~&+\frac{1}{3}\gamma^5K_5(2\gamma)\Big]\Big\},
\end{eqnarray}
Where $\phi=\frac{2\pi}{\lambda}w$ measures the deviation of the perpendicular projection of the sphere from a local hump on the surface.
Here $J_n(\gamma)$ , $I_n(\gamma)$ and $K_n(\gamma)$ are Bessel's function of the order $n$ \cite{GradRyzh}.

We can investigate the effects of the roughness in two extreme limits of long or short wavelength 
deformations. For the case of very short wave length deformations, $\lambda\ll H$ ($\gamma\gg1$) one can see that the roughness has 
no net 
effects on the self mobility of a spherical particle moving very far from the wall (up to first order of $\varepsilon$). 
In this case the mobility tensor is effectively given 
by the mobility tensor of a sphere located very far from a flat wall. 
The explicit form of the mobility tensor for this case is given by 
Eq. (\ref{selfmobility}). The back flows, scattered from the humps and deeps of the wall have cancelled out the effects of each other 
and the overall averaged back flow looks like a back flow from a flat wall.

In the case of a very long wavelength roughness where $\lambda\gg H$ ($\gamma\ll 1$), we proceed and obtain the second order corrections 
(in terms of $\varepsilon$) to the self mobility components. Now we can expand this result around small $\gamma$,  to reach the following expressions for the different elements of the self mobility tensor:
\begin{eqnarray}
\mu_{xx}^{R}&\approx&\mu_{xx}^{F}+\mu_0\frac{9}{4}(\frac{ah_{0}^{2}}{H^3})\Big[\frac{1}{32}
(1+5\cos(2\phi))\nonumber\\
&~~~&+3\cos^2(\phi)\Big],\nonumber\\
\mu_{yy}^{R}&\approx&\mu_{yy}^{F}-\mu_0\Big[\frac{27}{64}(\frac{ah_0}{H^2})
\cos(\phi)\nonumber\\
&~~~&-\frac{9}{4}(\frac{ah_{0}^{2}}{H^3})\left(\frac{1}{32}(1+5\cos(2\phi))+3\cos^2(\phi)\right)\Big],\nonumber\\
\mu_{zz}^{R}&\approx&\mu_{zz}^{F}+\mu_0\frac{27}{32}(\frac{ah_{0}^{2}}{H^3})
(1+\cos(2\phi)).
\end{eqnarray}

As one can see, the effects due to the roughness, enhances the self mobility tensor of a sphere in asymmetric way with respect to the 
in plane ($x$ and $y$) directions. We define the asymmetric parameter as the difference between the mobilities in the $x$ and $y$ directions:
\begin{equation}
\Delta\mu=\mu_{xx}^{R}-\mu_{yy}^{R}=\mu_0\left[\frac{27}{64}(\frac{ah_0}{H^2})
\cos(\phi)\right].
\end{equation}
\begin{figure}
\includegraphics[width=0.85\columnwidth]{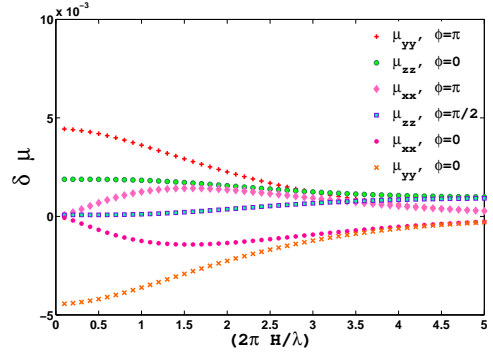}
\caption{(color online). Different components of self mobility tensor $\delta\mu_{ii}=(\mu^{R}_{ii}-\mu^{F}_{ii})/\mu_{ii}^{F}$ 
for a spherical particle moving adjacent to a non-smooth and no-slip wall are plotted as a functions of the particle's separation from the wall. 
Here the wave vector for surface roughness lies along the $x$  direction. All three components of the self mobility tensor are plotted for two special cases of motion above a hump or deep of the surface deformations. As one can see for $\phi=0$ (sphere move over a local hump), the roughness decreases the parallel components of self mobility while for $\phi=\pi$ (motion over a local deep), they increase. Parameter values for these graphs are: $h_0/H=0.1$, $a/H=0.1$.}
\label{fig3}
\end{figure}
\begin{figure*}
\includegraphics[width=0.97\columnwidth]{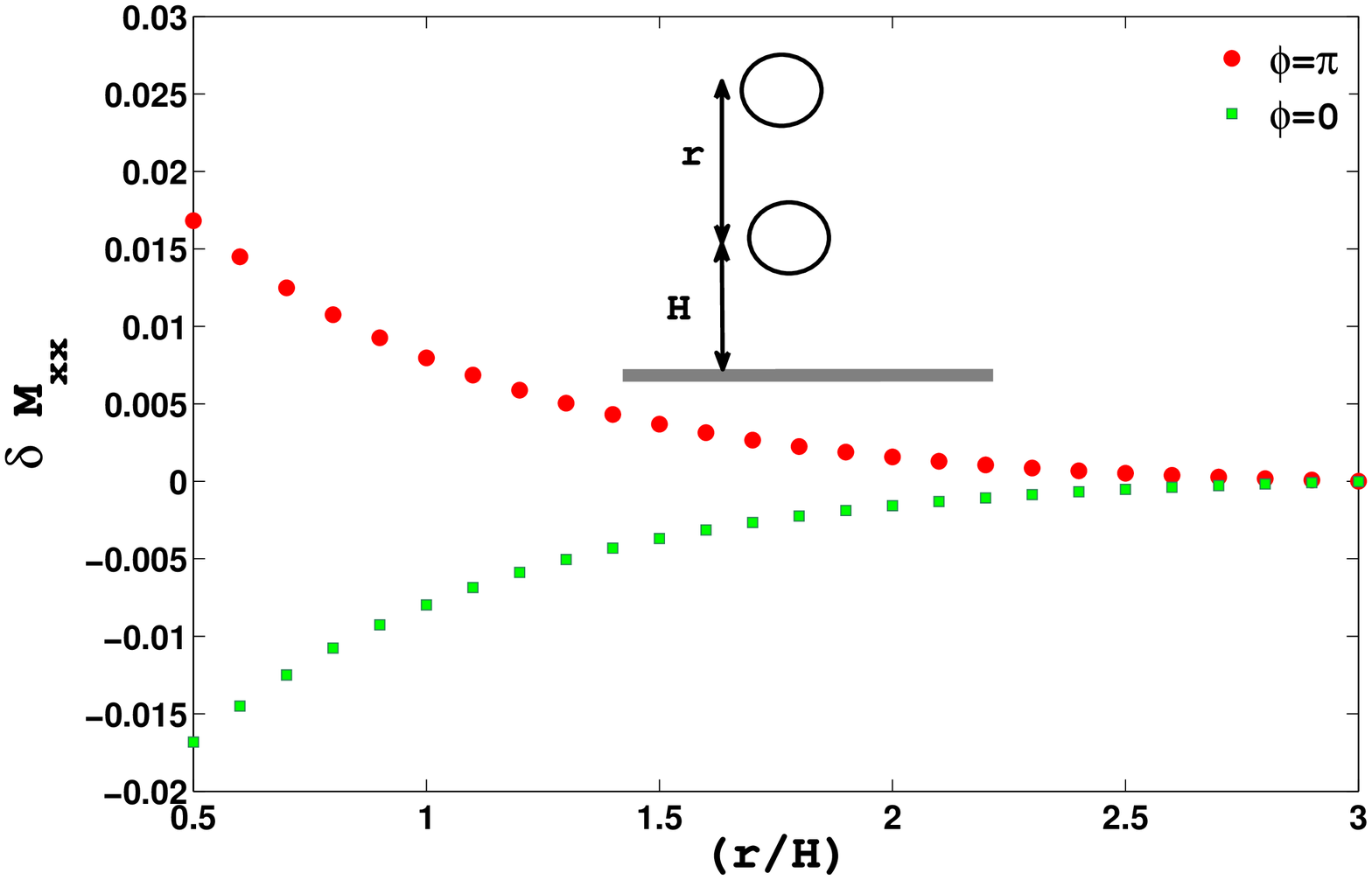}\includegraphics[width=0.97\columnwidth]{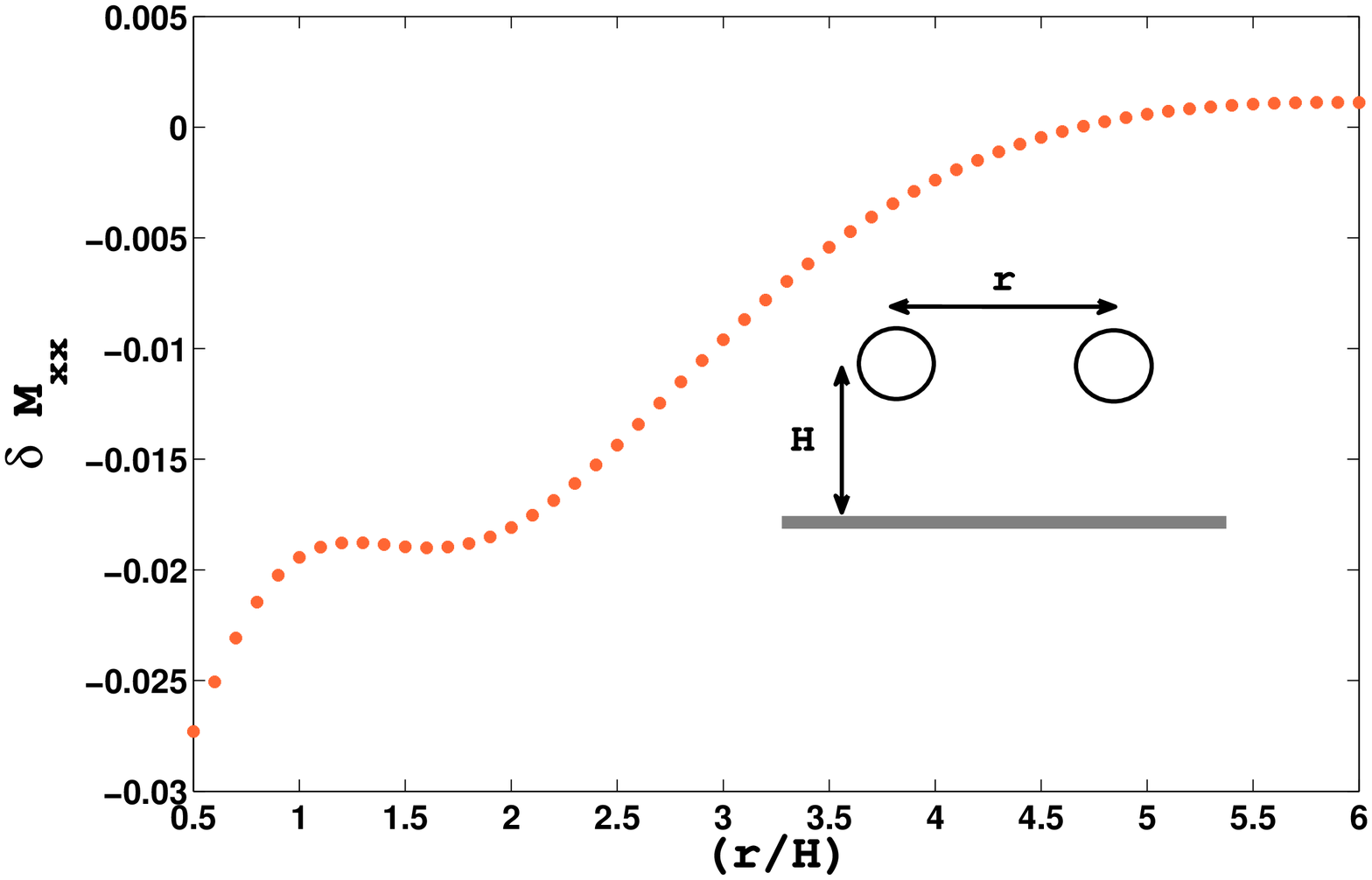}\\
\includegraphics[width=0.97\columnwidth]{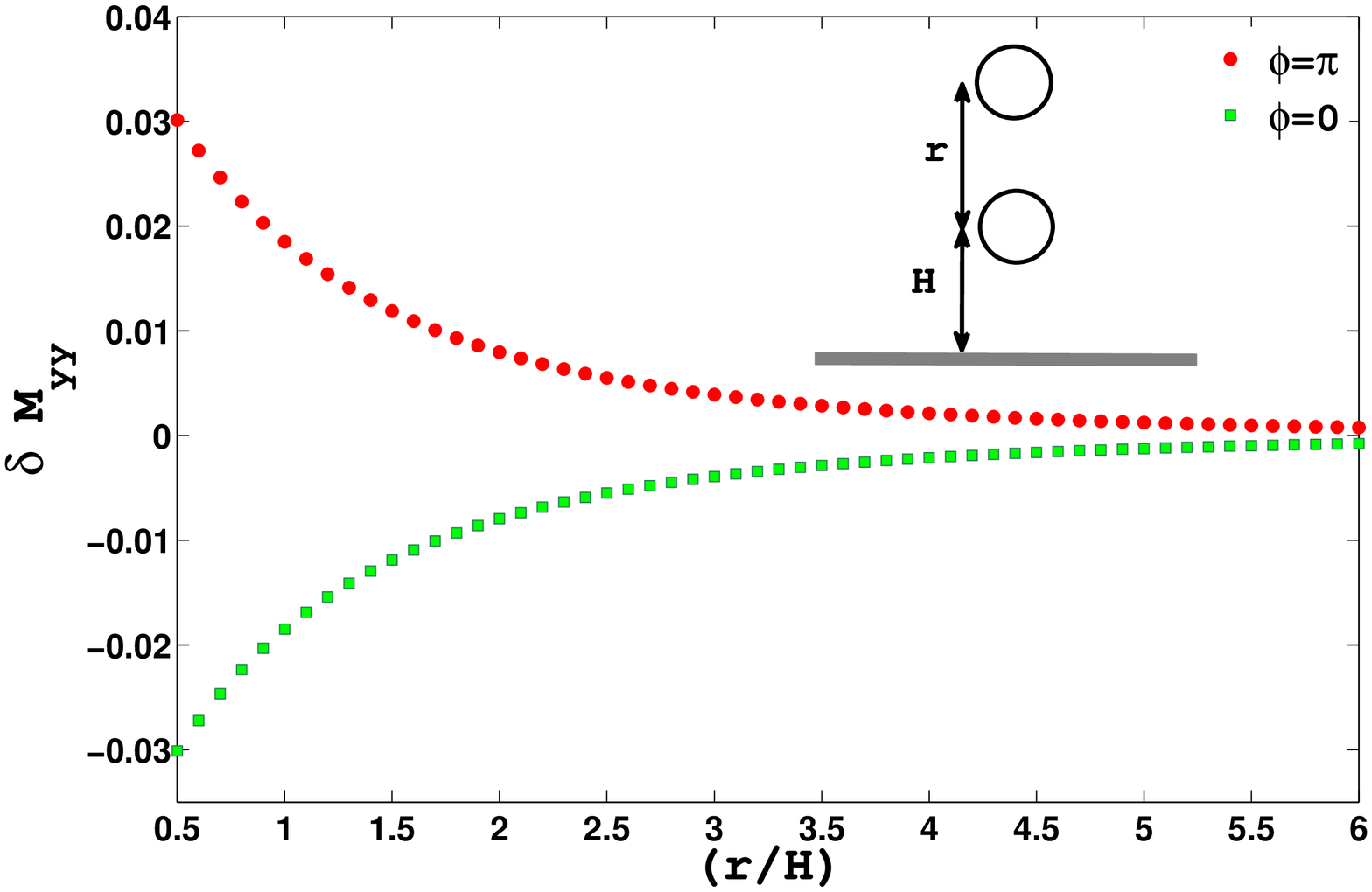}\includegraphics[width=0.97\columnwidth]{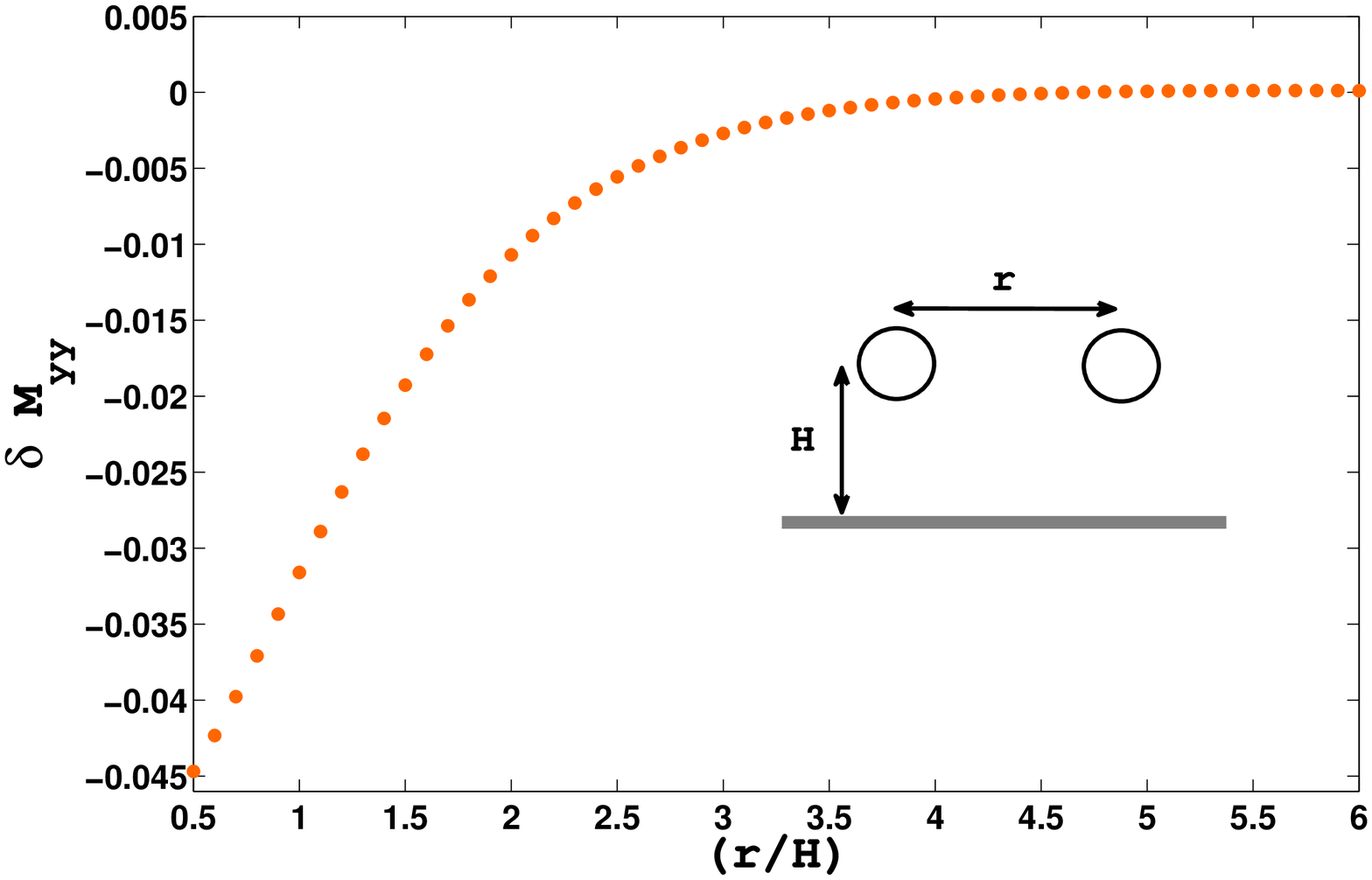}
\caption{(color online). Different components of the hydrodynamic interaction tensor $\delta M_{ii}=8\pi\eta H(M^{R}_{ii}-M^{F}_{ii})$, 
for two spherical particles moving parallel to the wall, are plotted as a functions of the particle's separation. Here the wave vector for surface roughness lies along the $x$  direction. At right graphs (up and down), the interaction is plotted for the 
case where the position of the first sphere is fixed at $\phi=0$ and we changed the position of the second sphere. There is a 
periodicity with the wavelength of the roughness.  As one can see in the case of left graphs, for $\phi=0$ (sphere move over a local hump), the roughness decreases the interaction strength while for $\phi=\pi$ (motion over a local deep), it increases. Parameter values for these 
graphs are: $\lambda/H=10$, $h_0/H=0.1$, $a/H=0.1$.}
\label{fig4}
\end{figure*}
\begin{figure}
\includegraphics[width=0.85\columnwidth]{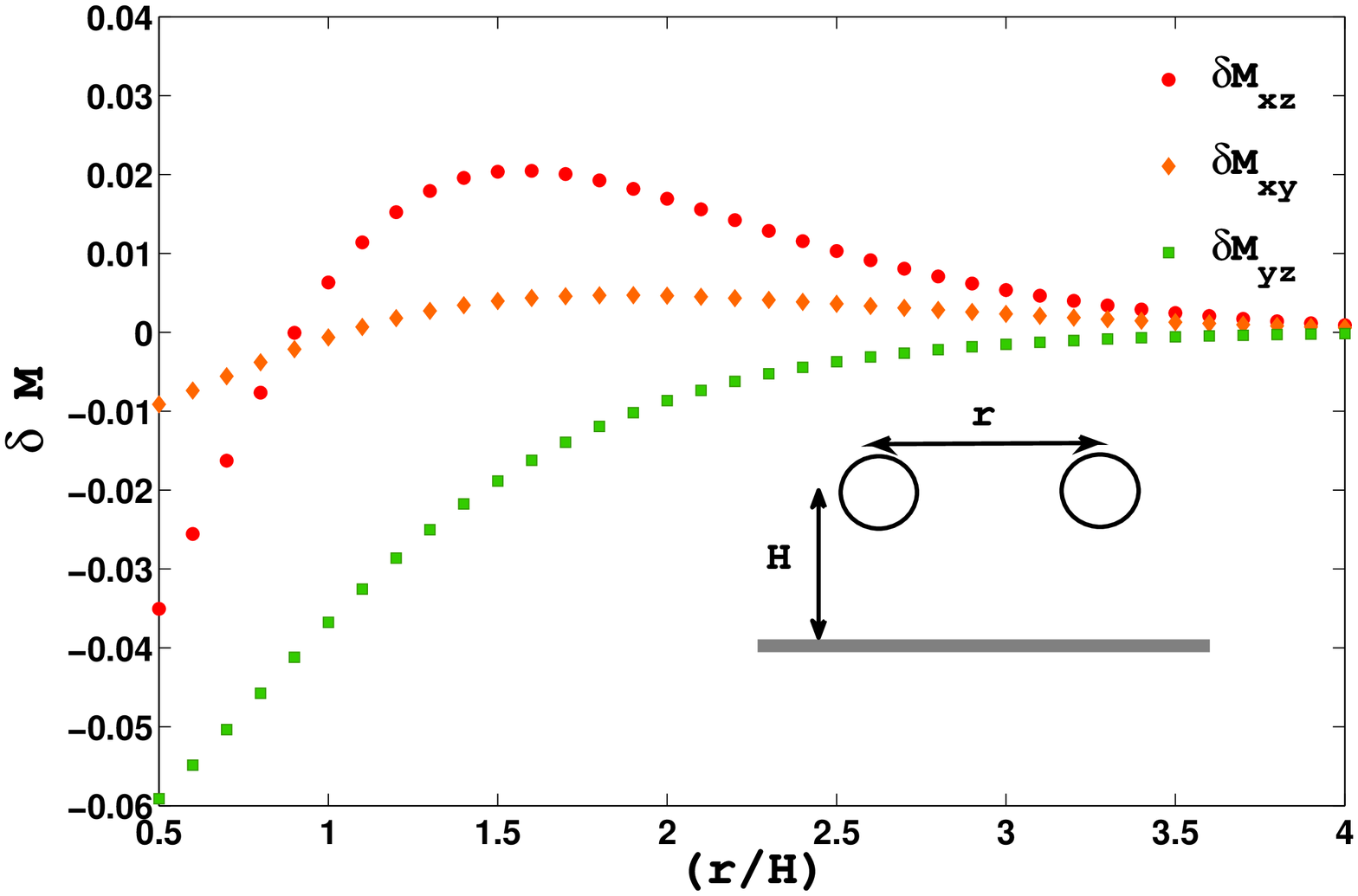}
\caption{(color online). Off diagonal components of the hydrodynamic interaction tensor $\delta M_{ij}=8\pi\eta H(M^{R}_{ij}-M^{F}_{ij})$, 
for two spherical particles, are plotted as a functions of the particle's separation. Here the wave vector for surface roughness lies along the $x$  direction. Parameter values for these graphs are: $\lambda/H=10$, $h_0/H=0.1$, $a/H=0.1$.}
\label{fig5}
\end{figure}

The sign of the asymmetric parameter depends on the local position of the sphere. 
In terms of the back flow scattered from the wall and in the limit of long wave 
length deformations, it is expected that the nearest hump or deep will have the dominant contributions on the particle motion. 
Over a local hump ($0<\phi<\pi/2$), 
$\mu_{xx}^{R}>\mu_{yy}^{R}$, while over a local deep ($\pi/2<\phi<3\pi/2$), $\mu_{xx}^{R}<\mu_{yy}^{R}$. Interestingly for a special points of 
$\phi=\pi/2,3\pi/2$, the mobility tensor is symmetric. 

To analyze the self mobility for intermediate $\gamma$ ($H\sim\lambda$), we have presented numerical plots in Fig. \ref{fig3}. As one can see for 
$H\sim\lambda$, the change in the mobility of a sphere mediated by roughness changes its sign for motion above a peak or above a valley.

To demonstrate the effects of roughness on the two-body interactions, we have presented in Fig. \ref{fig4} and Fig. \ref{fig5}, numerical results 
for diagonal and off diagonal elements of hydrodynamic interactions. 
In these examples we have studied the hydrodynamic interactions for some different cases.
In Fig. \ref{fig4} we have plotted $M_{xx}(r)$ (up) and $M_{yy}(r)$ (down). 
In Fig. \ref{fig4}, right part (up and down), we have plotted the interactions of two spheres that have same vertical distance $H$ from the wall. 
As one can see, the roughness always decreases the strength of $M_{xx}(r)$ and $M_{yy}(r)$. There is a very weak periodicity with the wavelength of the roughness.
In the left part of Fig. \ref{fig4} (up and down), we have plotted the hydrodynamic 
interaction as a function of vertical separation of two spherical particles. The results for this case,  
depend on the local position of the spheres with respect to the surface roughness. As one can see for $\phi=0$ 
(spheres move over a local hump), the roughness decrease the 
interaction strength while for $\phi=\pi$ (motion over a local deep), it increases. 
As another example, some of  off diagonal components of the two body hydrodynamic interactions are plotted in Fig. \ref{fig5}. There is a very weak periodicity with wavelength $\lambda$, that is not clearly seen in the scale of this graph.

\section{concluding remarks}
In this article we have considered the influence of a rough, rigid and no-slip boundary wall on the hydrodynamic interactions 
of spherical particles. We have studied a regular sinusoidal roughness pattern with very small amplitude roughness on a flat plane. 
For simplicity we have studied a simple wave with a single wave vector along $x$ direction.
Taking into account the wall effects by applying the no-slip boundary condition by standard perturbation technique, we have 
calculated one and 
two-body hydrodynamic effects. For a single and small radius sphere moving in the presence of a rough wall, we show that, 
the different elements of  self mobility tensor changes in asymmetric way. Motion along the wave vector 
of surface roughness is different from the other in-plane direction. When a spherical colloid suspended near a rough and 
no-slip wall, the hydrodynamic drag force depends on the local position of the sphere. Roughness will produce different 
contributions for motion on a local hump or a local deep of the wall. This kind of behavior in two-body hydrodynamic 
interaction is also seen by numerical investigations of different components of the hydrodynamic interactions.

We note that in the current formulation of the problem, we have made some approximations which need to be dealt with carefully. Caution is
needed in applying the results since there are many length scales
in the problem: $a$, $h_0$, $\lambda$ and $H$. First, the Faxen's formula has allowed us to  treat the  sphere's size in a series expansion in powers of $\epsilon_{1}=(a/H)$. Second approximation is related to the slowly varying roughness of the wall  and consequently to a series expansion of the results in powers  of $\epsilon_{2}=(h_0/H)$. The series expansion for a typical component of the hydrodynamic interactions, for example the self mobility, have the following structure:
\begin{eqnarray}
\Delta \mu/\mu_0=\epsilon_{1}\times&\Big\{&\left[ f_{11}(\gamma)\epsilon_{2}+f_{12}(\gamma)\epsilon_{2}^2+{\cal O}(\epsilon_{2}^3)\right]\nonumber\\
&+&\left[ f_{21}(\gamma)\epsilon_{2}+f_{22}(\gamma)\epsilon_{2}^2+{\cal O}(\epsilon_{2}^3)\right]\epsilon_{1}\nonumber\\
&+&{\cal O}(\epsilon_{1}^2)\Big\},
\end{eqnarray}
where we have already defined $\gamma=2\pi(H/\lambda)$. The results are valid for any $\gamma$, however one should note that, 
the convergence of the above series expansion in the limit of very small $\gamma$ ($\gamma\ll 1$), is constrained to the condition 
$(h_0/\lambda)\ll 1$. By small roughness assumption, one expects that this criterion is satisfied as well.

In conclusion, we have developed a systematic way to evaluate the perturbation effects of a rough wall in the hydrodynamical properties of small spherical particles. The results of current work can be used in many directions related to the colloidal problems in confined flows, where the roughness 
is an ignorable characteristics of most walls. 
The influence of wall roughness on the thermal diffusion of colloidal particles  is an interesting issue that we are 
considering. Inspired by the ensemble of
low Reynolds swimmers, we are also investigating the motion of low Reynolds self propellers adjacent to a rough wall.
\acknowledgments
We thank Ramin Golestanian for stimulating discussions and also for reading the manuscript. 


\begin{thebibliography}{99}
\bibitem{infinite} {As a simple prescription for regularization, in calculating the components of self mobility, 
we replace all divergent terms with the self mobility of a sphere $1/(6\pi\eta a)$, that is also divergent for a very small sphere.}
\end{thebibliography}
\end{document}